\documentclass{article}
\usepackage{spconf,amsmath,graphicx,hyperref}
\usepackage{booktabs}
\title{Advancing Speaker Based Vocal Effort Classification with WavLM and Data Augmentation in Naturalistic Non-Calibrated Speech Recordings\thanks{This work was supported in part by the University of Texas at Dallas (UTDallas) from the Distinguished University Chair in Telecommunications Engineering held by J.H.L. Hansen.}}

\name{Zahra Omidi, John H. L. Hansen}
\address{Author Affiliation(s)}
\address{Center for Robust Speech Systems, \\
The University of Texas at Dallas, USA}
\begin{document}

\maketitle

\begin{abstract}
The variations in vocal effort range (e.g. whisper, soft, neutral, loud, shout) alter production and speech acoustics, reducing intelligibility and limiting the robustness of any subsequent speech technology. Classification is challenging since effort lies on a continuum, adjacent categories are easily confused, and labeled data remain scarce. Prior SSL approaches with wav2vec2, HuBERT, and AST improve performance on the AVID corpus but still suffer from boundary errors. In this study, we introduce WavLM for the first time in vocal effort classification and benchmark it against wav2vec2 and HuBERT. To address data scarcity, we conduct a systematic study of augmentation strategies, covering RIR convolution, additive noise, time masking, speed perturbation, band-limiting, MixUp, and CutMix. Augmentation consistently improves WavLM, with gains ranging from +0.6\% to +1.8\% absolute. We further propose Gaussian-neighbor soft labels, which further reduce near-boundary confusions by modeling the vocal effort continuum. Our best system, WavLM-BASE with gradual unfreezing, augmentation, and Gaussian-neighbor soft labels, achieves 78.2\% mean accuracy, establishing a new state-of-the-art on AVID.
\end{abstract}


\section{Introduction}
Vocal effort, ranging from whisper, soft, neutral, loud, and shouted speech, alters speech production and acoustic structure and impacts both intelligibility and performance of speech systems. Prior studies have examined this continuum, showing spectral and energy shifts that degrade robustness in speaker identification \cite{Zhang2007SpeechMode, Hughes2023ForensicEffort, Prieto2022VocalEffortSV}. 
Related work has furhter investigated whispered speech in automatic speech recognition and speaker recognition settings, highlighting similar challenges under reduced vocal effort \cite{GhaffarzadeganBorilHansen2017WhisperASR,KellyHansen2021LombardWhisper}.
While such findings underline the importance of modeling vocal effort across the full range, only the UT-VE-I \cite{zhang2018advancements, ZhangHansen2015WhisperIsland} corpus provides recordings spanning whisper through shout. Among more recent resources, the AVID corpus \cite{Alku2024AVID} offers another alternative, covering soft to very loud speech with a larger pool of labeled data, and thus serves as our dataset for supervised vocal effort classification.

Unlike conventional ASR, vocal effort classification requires separating subtle acoustic shifts between adjacent categories while remaining robust to channel and environmental variability. This challenge is further compounded by the limited size of available corpora. 
Several feature types have been investigated for this task, including spectrogram representations, speaker embeddings, and wavelet scattering features \cite{Kodali2025WSN} with embeddings extracted from self-supervised learning (SSL) models achieving the best performance. Fine-tuning SSL encoders, specifically Wav2Vec2 \cite{Baevski2020Wav2Vec2}, HuBERT \cite{Hsu2021HuBERT}, and Audio Spectrogram Transformer (AST) \cite{Gong2021AST}, yield clear gains over traditional features, although confusions between neighboring levels remain \cite{Kodali2024FineTuning}. Earlier work also examined embeddings from Wav2Vec2 and Whisper without fine-tuning \cite{Kodali2023Whisper}. These studies establish SSL fine-tuning for vocal effort classification, while unresolved challenges exist at class boundaries.

Data augmentation has proven effective across domains, including image/text recognition \cite{Omidi2024EndToEndHDSR} and speech tasks such as SER and robust ASR, with techniques ranging from RIR convolution and additive noise to SpecAugment masking \cite{Park2019SpecAugment}. Mix-based methods, including MixUp and CutMix, have also shown potential for continuous-label speech tasks \cite{Kim2021SpecMix}. In parallel, label smoothing and other soft-label training strategies have reduced overconfidence and improved generalization in speech and vision tasks \cite{Muller2019LabelSmoothing,Xu2020SoftLabel}.Here, we explore different augmentation and soft-labeling methods to address data scarcity and boundary confusions that characterize Vocal Effort Identification (VE-ID).

Soft labels act as a regularizer by replacing one-hot targets with probability distributions, thereby reducing overconfidence and improving generalization. Here, we use label smoothing \cite{Muller2019LabelSmoothing} and mix-based augmentations MixUp and CutMix \cite{Zhang2018Mixup,Yun2019CutMix}, together with Gaussian-neighbor soft labels tailored to vocal effort categories. These approaches mitigate near-boundary errors and yield more robust models under limited data conditions compared to hard-label training.

In this study, we fine-tune WavLM \cite{Chen2022WavLM}, and conduct a systematic comparison of three SSL backbones (wav2vec2 \cite{Baevski2020Wav2Vec2}, HuBERT \cite{Hsu2021HuBERT}, and WavLM), identifying WavLM-BASE as the strongest model. Next, we study waveform-level augmentations (RIR convolution, additive noise, time masking \cite{Park2019SpecAugment}, speed perturbation, band-limiting) along with mix-based augmentation methods (MixUp, CutMix \cite{Kim2021SpecMix}). Finally, to further reduce VE-ID boundary errors, we adopt Gaussian-neighbor soft labels with class-specific variance and skew \cite{Muller2019LabelSmoothing,Xu2020SoftLabel}. Our contributions are threefold: (1) benchmark SSL encoders for vocal effort classification and show that WavLM-BASE, combined with gradual unfreezing of transformer layers, provides the most effective backbone; (2) present the first systematic study of augmentation strategies for vocal effort classification, with performance gains across VE scale; (3) propose Gaussian-neighbor soft labels tailored to VE categories, which further improve accuracy and reduce confusions between adjacent levels.

\section{Methodology}
\vspace{-0.1in}

\subsection{Dataset}
\vspace{-0.05in}
All experiments are performed using AVID corpus \cite{Alku2024AVID}, originally designed for vocal effort. 
The corpus contains recordings from 50 English speakers (25 male, 25 female), each asked to read prompted sentences at four instructed vocal intensity levels: \textit{soft}, \textit{normal}, \textit{loud}, and \textit{very loud}. 
Recordings were made in a laboratory setting using a close-talking microphone, with repeated utterances across different sentences and speakers. 
Thus, AVID has 10,000 labeled utterances evenly distributed across four effort categories.  

Following prior work \cite{Kodali2024FineTuning}, we use the non-calibrated version of the corpus, where absolute amplitude and sound pressure level (SPL) information removed by waveform normalization. This condition makes the task more challenging since classification must rely on relative spectral-temporal cues rather than overall amplitude.  

Evaluation is carried out using group $K$-fold cross-validation with $K=10$. Accuracy is reported as mean $\pm$ standard deviation across folds, consistent with previous studies on the AVID corpus.

\vspace{-0.05in}
\subsection{Models}
\vspace{-0.05in}
We evaluate three self-supervised learning (SSL) encoders widely adopted in speech processing: Wav2Vec2-Base \cite{Baevski2020Wav2Vec2}, HuBERT-Base \cite{Hsu2021HuBERT}, and WavLM-Base \cite{Chen2022WavLM}. 
All are transformer-based pre-trained on large-scale speech corpora with masked prediction objectives, and shown to learn robust acoustic and prosodic representations.

We restrict our experiments to \textit{Base} configurations of each model. While prior work on VE-ID also considered the Large variants \cite{Kodali2024FineTuning}, accuracy improvements were marginal. Given the limited size of labeled data, Base models offer a balance between performance and efficiency, and are therefore the focus of this study.

Each model is fine-tuned end-to-end on the VE-ID task. A lightweight classification head consisting of two fully connected layers with ReLU activation and dropout is attached on top of the final transformer layer. The head outputs posterior probabilities across the four vocal intensity categories (\textit{soft}, \textit{normal}, \textit{loud}, \textit{very loud}). 
Training is performed with cross-entropy loss for hard labels, and KL-divergence against soft distributions using sub-label supervision. 

\vspace{-0.1in}
\subsection{Data Augmentation}
\vspace{-0.05in}
Given limited amount of labeled data for VE-ID, we apply a broad set of waveform-level augmentations to increase variability and improve robustness. All methods are implemented directly in the time domain to preserve information relevant to effort cues.
\begin{figure}[t]
  \centering
  \includegraphics[width=0.8\linewidth]{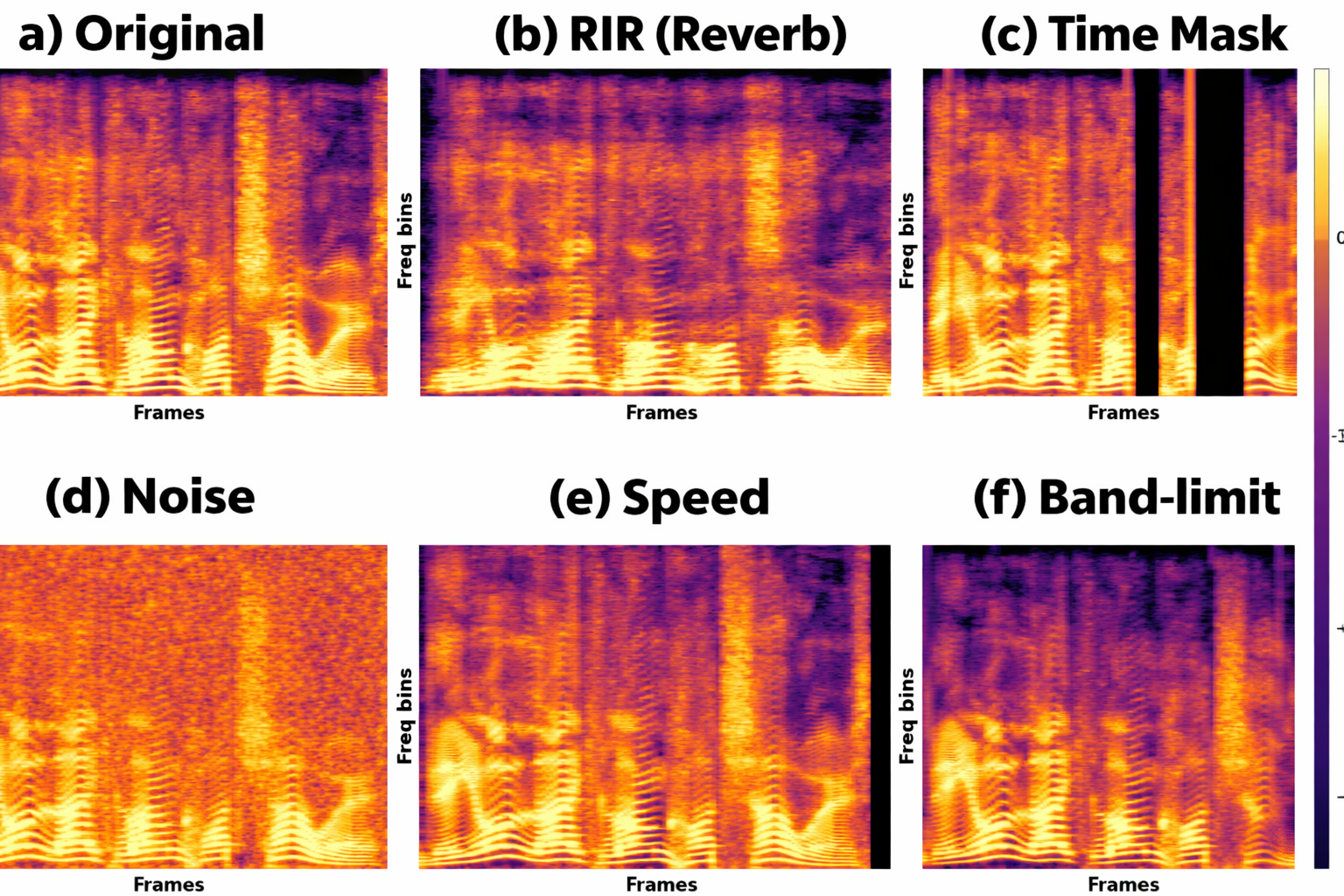}
  \caption{Log-mel spectrograms of an example utterance under different data augmentations: (a) original, (b) RIR convolution (reverberation), (c) time masking, (d) additive noise, (e) speed perturbation, and (f) band-limiting.}
  \label{fig:augment_specs}
  \vspace{-2pt}
\end{figure}

The augmentations include: \textbf{RIR convolution} - using real room impulse responses, trimmed to early reflections and RMS-normalized to simulate reverberation; \textbf{Additive noise} - adding Gaussian noise at 20--30~dB SNR; \textbf{Time masking} - replacing random segments of up to 10\% of the waveform with silence; \textbf{Speed perturbation} - resampling with factors between 0.98--1.2 while preserving effort cues by avoiding RMS matching; and \textbf{Band-limiting} - applying low/high-pass filtering with a dry/wet mix to mimic diverse channel effects. We also employ mix-based strategies: \textbf{MixUp} - which linearly interpolates two utterances and their labels, and \textbf{CutMix} - which replaces contiguous segments between utterances with labels mixed proportionally. All augmentations are sampled randomly during training, with parameters redrawn at each epoch, and only one is applied per utterance unless otherwise stated.

\begin{figure}[t]
    \centering
    \includegraphics[width=0.75\linewidth]{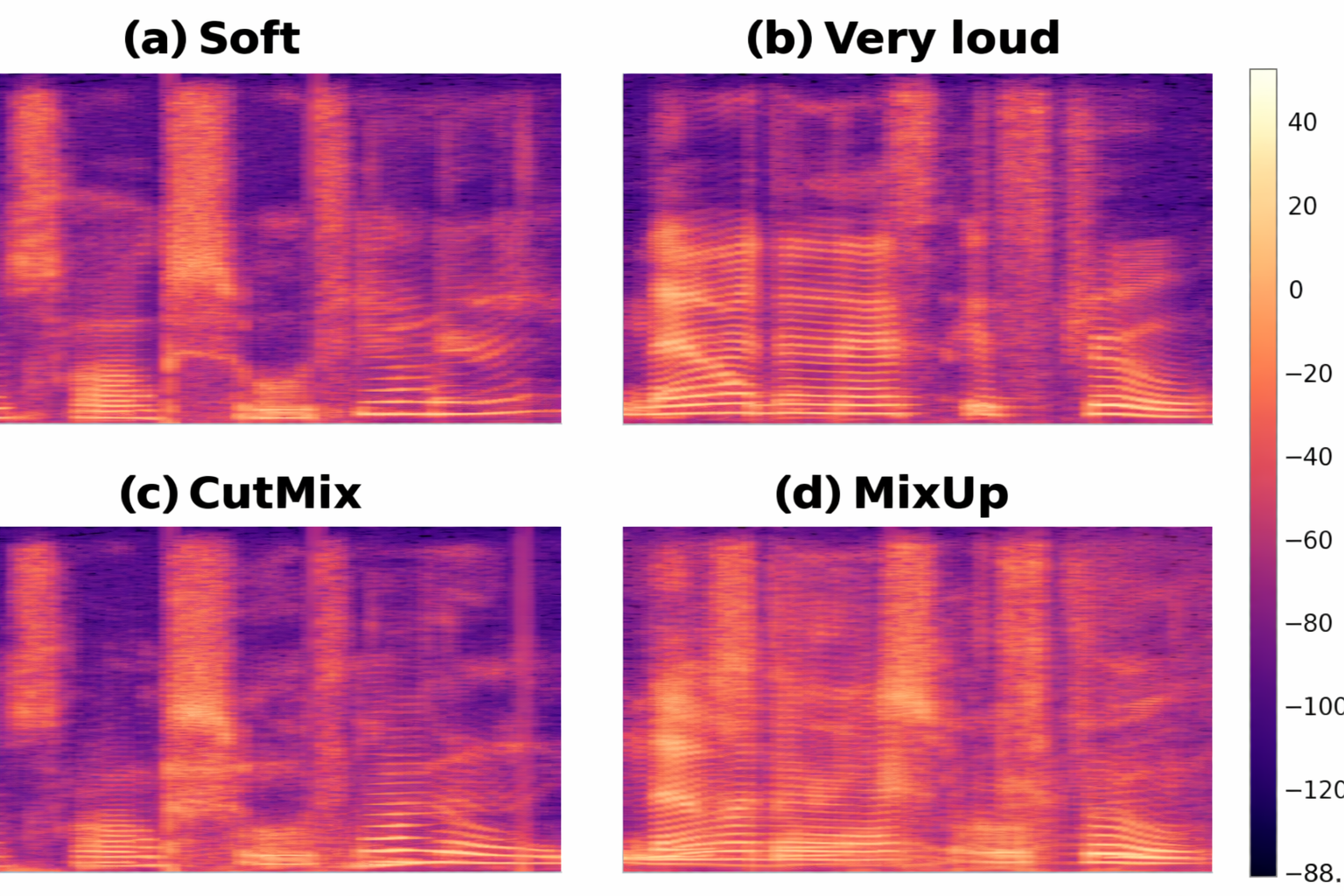}
    \caption{Mix-based data augmentations. (a–b) Source utterances (\emph{Soft}, \emph{Very loud}). (c) \emph{CutMix}, where a contiguous segment is replaced. (d) \emph{MixUp}, which performs a global interpolation between utterances. All spectrograms are shown using a shared dB scale.}
    \label{fig:mix_augs}
\end{figure}

Figure~\ref{fig:augment_specs} illustrates their effects in the spectrogram domain: reverberation smears energy across time, noise fills high-frequency regions, speed perturbation compresses or expands harmonics, time masking removes short spans, band-limiting suppresses frequency bands. Figure~\ref{fig:mix_augs} further contrasts MixUp and CutMix: MixUp produces global blends across the entire signal, while CutMix introduces sharp, localized segment-level transitions. Together, these strategies expand the training distribution beyond discrete class boundaries by generating intermediate and composite examples.

\vspace{-0.05in}
\subsection{Soft-Label Regularization}
\vspace{-0.05in}

Hard one-hot labels assume categorical boundaries, but VE is perceptually continuous, and annotator disagreement is concentrated near neighboring VE. Soft-label training regularizes learning by replacing one-hot targets with probability distributions, reducing overconfidence. Here, we investigate three forms of soft labels:

\textbf{Label smoothing:} We distribute a small portion of the probability mass uniformly across all classes, discouraging overconfidence and improving calibration.

\textbf{Gaussian-neighbor soft labels:} To explicitly model proximity between VE levels, we replace one-hot targets with Gaussian-smoothed distributions centered on the ground-truth class. This encourages the model to recognize ambiguity near category boundaries.

\textbf{Mix-based soft labels:} We introduce a variant of MixUp \cite{Zhang2018Mixup} and CutMix \cite{Yun2019CutMix} where, instead of interpolating one-hot labels, we interpolate Gaussian-neighbor soft labels. This way, the mixed targets follow the mixing weights and the ordinal structure of VE. 

\begin{figure}[t]
    \centering
    \includegraphics[width=0.75\linewidth]{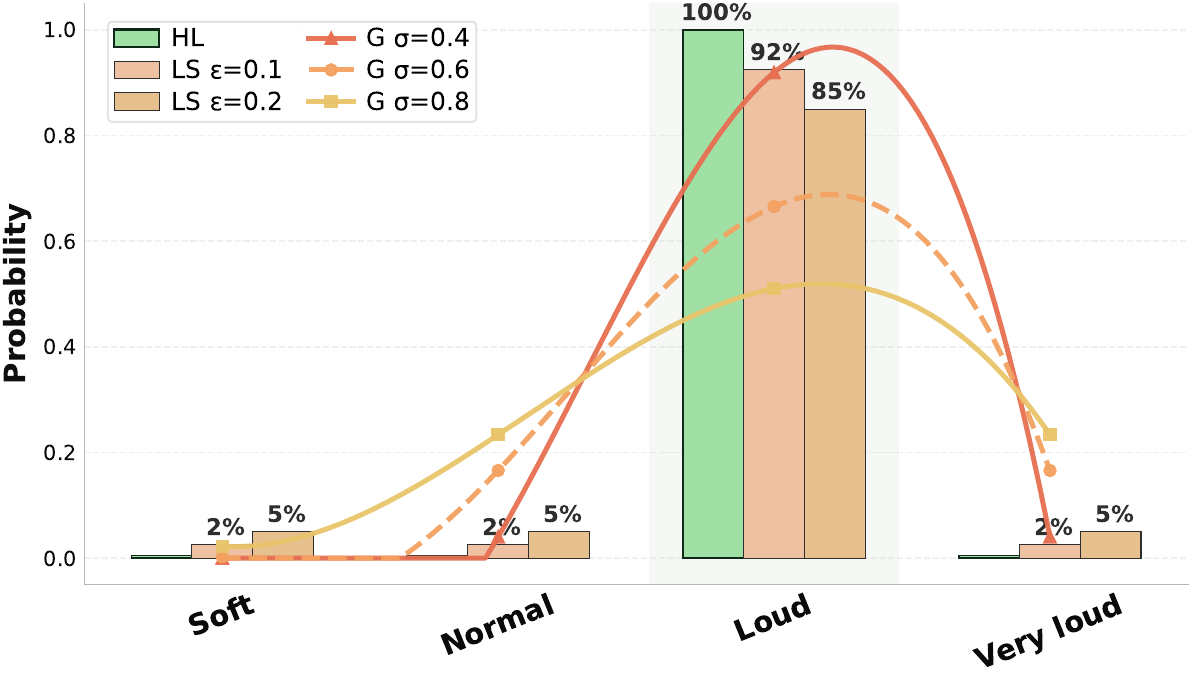}
    \caption{Soft-label strategies for the \emph{loud} category (HL: hard label; LS: label smoothing; G: Gaussian-neighbor).}
    \label{fig:softlabels}
\end{figure}

Figure~\ref{fig:softlabels} illustrates the effect of different Soft-label strategies. 
Hard labels assign full probability to a single class, while label smoothing redistributes a fixed portion of the mass across all categories. 
In contrast, Gaussian-neighbor formulation concentrates probability on neighboring effort levels, with variance $\sigma$ controlling how broadly the distribution spreads. 
This more closely reflects the perceptual continuum of VE, where confusions typically occur between adjacent levels, and provides a stronger regularizer than uniform smoothing.
All soft-label training is implemented with KL-divergence between predictions and soft targets. 

\section{Experiments}
\subsection{Experimental Setup}

All experiments are conducted on the AVID corpus \cite{Alku2024AVID} under the non-calibrated condition, using the four instructed effort categories. Evaluation follows 10-fold group cross-validation, and results are reported as mean accuracy with standard deviation across folds. Models are fine-tuned end-to-end with Adam (encoder learning rate $5\times10^{-6}$, classifier $1\times10^{-4}$), batch size 32, and 15 epochs. All audio is resampled to 16~kHz and mean-variance normalized before input.

\label{sec:results}
    \vspace{-2pt}
    \begin{table}[h!]
    \centering
    \caption{Comparison of SSL models (Base) on AVID corpus. Mean Acc.\ (\%) $\pm$ Std (10-fold CV).}
    \begin{tabular}{lcc}
    \toprule
    \textbf{Model} & \textbf{Mean Accuracy $\%$} & \textbf{Std} \\
    \midrule
    Wav2Vec2-Base  & 67.58 & 1.50 \\
    HuBERT-Base    & 74.13 & 1.22 \\
    \textbf{WavLM-Base} & \textbf{75.24} & \textbf{1.47} \\
    \bottomrule
    \end{tabular}
    \label{tab:models}
    \vspace{-2pt}
\end{table}

\subsection{Results: Baselines vs WavLM}
As shown in Table~\ref{tab:models}, WavLM-Base yields the strongest performance among the evaluated base SSL encoders, outperforming wav2vec2-Base and HuBERT-Base by more than $1\%$ absolute. All three encoders have comparable model capacities (approximately 94–95M parameters), ensuring that the observed differences are not attributable to model size. For WavLM-Base, we employ a gradual unfreezing strategy, in which the SSL encoder is initially frozen and then progressively unfrozen by one layer per epoch to improve training stability during fine-tuning. This approach prevents early overfitting and lets deeper layers adapt more effectively to subtle vocal effort cues, leading to stable training and consistent gains. As a result, WavLM-Base serves as a strong and reliable backbone for all subsequent studies.

\begin{table}[h]
\centering
\caption{Ablation of augmentation methods on WavLM-Base. Mean Acc.\ (\%) $\pm$ Std (10-fold CV)
}
\begin{tabular}{lcc}
\toprule
\textbf{Augmentation method} & \textbf{Mean Acc.} & \textbf{Std} \\
\midrule
No augmentation    & 75.24 & 1.47 \\
\midrule
Additive Noise     & 75.86 & 1.74 \\
Band limit         & 76.54 & 1.42 \\
Speed perturbation & 76.63 & 1.27 \\
Time masking       & 76.79 & 1.34 \\
RIR convolution    & 76.93 & 1.46 \\
RIR + time + noise & 76.54 & \textbf{1.16} \\
\midrule
CutMix             & 76.91 & 1.48 \\
MixUp              & \textbf{77.00} & 1.52 \\
\bottomrule
\end{tabular}
\label{tab:aug_ablation}
\end{table}

\subsection{Effect of Augmentation}
Next, Table~\ref{tab:aug_ablation} shows all augmentation methods improve performance over no-augmentation baseline. Among the standard approaches, RIR convolution and time masking yield the strongest gains, confirming their effectiveness in simulating reverberant conditions and encouraging reliance on longer-term context. Additive noise offers only marginal improvement, suggesting that introducing channel noise does not help distinguish VE. Interestingly, combining RIR, time masking, and noise does not outperform single augmentations, likely because added noise dominates and reduces the benefit of the other perturbations. 

Mix-based methods provide the largest benefits. MixUp achieves the best overall accuracy, consistent with vocal effort production lies on a continuum: so interpolated samples help guide the model toward smoother decision boundaries. CutMix performs similar to the best standard augmentations but lags MixUp, potentially due to disruptive prosodic structure.  

Finally, standard deviations are low across all conditions (1.16–1.74), confirming that improvements are consistent across folds. 

\begin{table}[h]
  \centering
  \caption{Comparison of mix-based augmentation and label-softening strategies on WavLM-Base. 
  GN denotes Gaussian-Neighbor soft labels. Mean Acc.\ (\%) $\pm$ Std (10-fold CV)}
  \label{tab:mix_soft}
  \begin{tabular}{lcc}
    \toprule
    Method & Mean Acc. (\%) & Std \\
    \midrule
    Hard labels           & 75.24 & 1.47 \\
    Label smoothing       & 76.95 & 1.45 \\
    GN (Soft labels)      & \textbf{77.32} & 1.46 \\
    \midrule
    MixUp                      & 77.00 & 1.52 \\
    MixUp + GN        & 77.27 & 1.47 \\
    CutMix                     & 76.91 & 1.48 \\
    CutMix + GN       & 77.18 & 1.32 \\
    \midrule
    MixUp ($\alpha=0.8$) + GN & 76.85 & 1.29 \\
    MixUp ($\alpha=0.6$) + GN  & \textbf{78.22} & \textbf{1.18} \\
    \bottomrule
  \end{tabular}
\end{table}
\vspace{-0.1in}
\subsection{Effect of Soft Labels}
Table~\ref{tab:mix_soft} compares hard labels, label smoothing, and Gaussian-neighbor soft labels, with and without MixUp and CutMix. Label smoothing improves over baseline by +1.7\% absolute, while Gaussian-neighbor soft labels provide a larger gain. These results show that modeling proximity between effort levels yields more benefit than uniform smoothing, reducing boundary errors and aligning better with the continuum nature of vocal effort. 

\begin{figure}[t]
    \centering
    \includegraphics[width=\linewidth,]{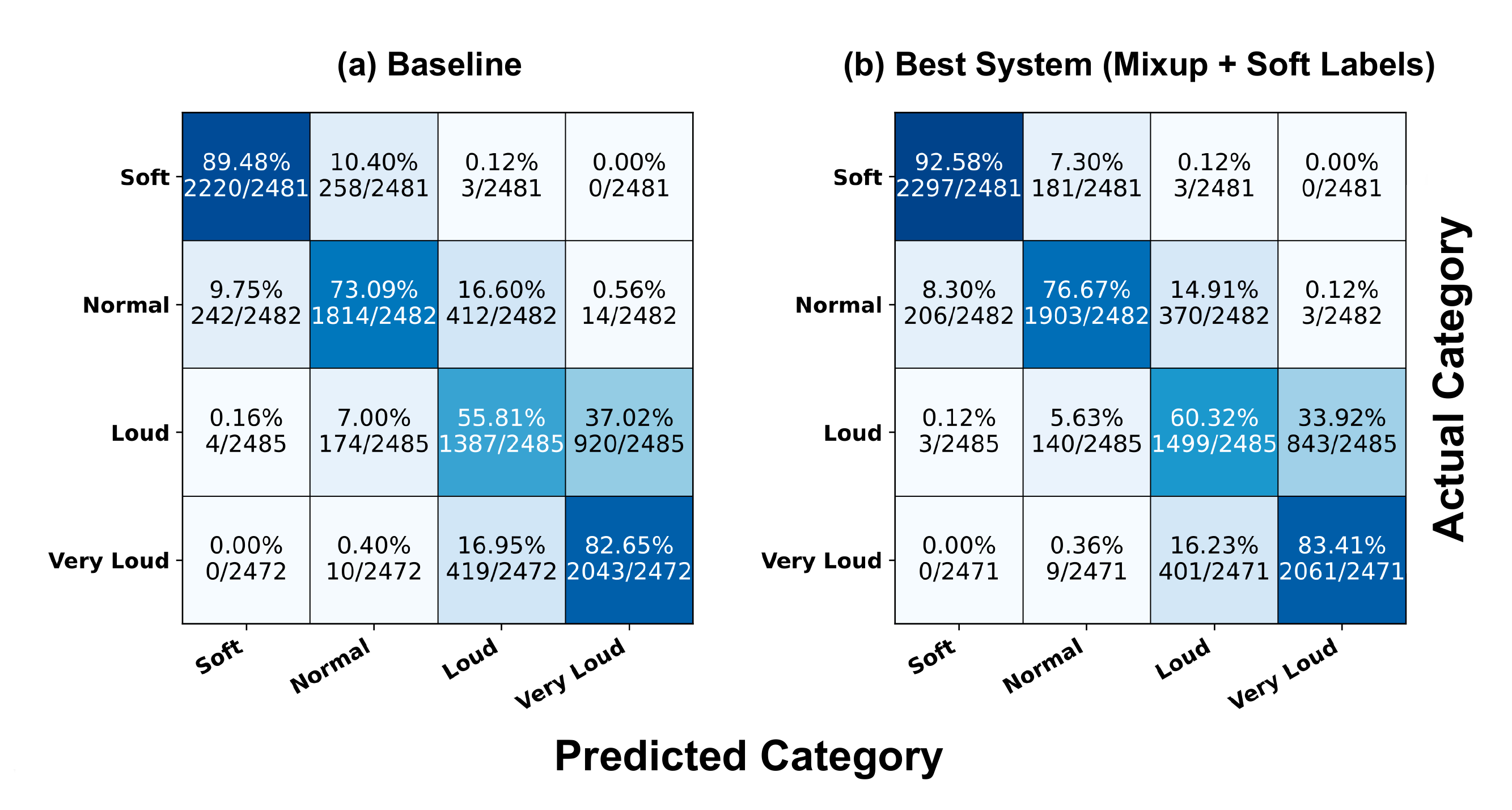}
    \caption{Confusion matrices for VE-ID. (a) Baseline WavLM-Base with hard labels shows heavy confusions between adjacent levels (e.g., \emph{loud} vs. \emph{very loud}). (b) Best system (MixUp $\alpha=0.6$ + Gaussian neighbor soft labels) reduces boundary errors and achieves balanced class distribution.}
    \label{fig:confmat}
\end{figure}

Finally, adjusting MixUp sharpness $\alpha=0.6$ achieves the best overall performance with the lowest variance. By comparison, $\alpha=0.8$ yields 76.85\%, suggesting over aggressive mixing can blur category boundaries. Overall, Gaussian-neighbor soft labels consistently improve over hard labels and uniform smoothing, and their integration with MixUp provides the strongest and most stable results.
For completeness, Fig.~\ref{fig:confmat} shows confusion matrices comparing baseline and best system (MixUp + Gaussian soft labels). 

\vspace{-0.1in}
\section{Conclusion}
This study has considered advancements in vocal effort classification for naturalistic speech data. This is the first study of WavLM for vocal effort classification and showed that it outperforms wav2vec2 and HuBERT by over 7\% and 1\% absolute, respectively. A systematic evaluation of augmentation strategies demonstrated consistent gains of +0.6--1.8\% absolute, with MixUp and RIR providing the strongest improvements. We further introduced Gaussian-neighbor soft labels, which improved accuracy by +2.1\% over hard labels and, when combined with MixUp at $\alpha=0.6$, achieved an overall 78.22\% mean accuracy with the lowest variance to date on the AVID vocal effort corpus. Gradual unfreezing proved essential for stable fine-tuning of WavLM in this limited-data setting. Together, these results establish WavLM with augmentation with Gaussian-neighbor soft labels as a new effective model for vocal effort classification for naturalistic data. Future work will explore massive vocal effort classification on naturalistic team-based communications such as Fearless Steps APOLLO \cite{Hansen2024FearlessSteps}.

\vspace{-0.1in}
\bibliographystyle{IEEEbib}
\bibliography{strings,refs}

\begin{thebibliography}{10}

\bibitem{Zhang2007SpeechMode}
C.~Zhang and J.~H.~L. Hansen,
\newblock ``Analysis and classification of speech mode: Whispered through shouted,''
\newblock in {\em Proc. Interspeech}, 2007, pp. 2289--2292.

\bibitem{Hughes2023ForensicEffort}
V.~Hughes, A.~Eriksson, and A.~Kachkovskiy,
\newblock ``Forensic speaker recognition under vocal effort variation,''
\newblock in {\em Proc. Interspeech}, 2023, pp. 2333--2337.

\bibitem{Prieto2022VocalEffortSV}
A.~Prieto, A.~Miguel, and E.~Lleida,
\newblock ``Speaker verification under vocal effort variation: Compensation approaches,''
\newblock in {\em Proc. Interspeech}, 2022, pp. 2913--2917.

\bibitem{GhaffarzadeganBorilHansen2017WhisperASR}
S.~Ghaffarzadegan, H.~Boril, and J.~H.~L. Hansen,
\newblock ``Deep neural network training for whisper speech recognition using small databases and generative model sampling,''
\newblock {\em International Journal of Speech Technology}, vol. 20, no. 4, pp. 1063--1075, Dec. 2017.

\bibitem{KellyHansen2021LombardWhisper}
F.~Kelly and J.~H.~L. Hansen,
\newblock ``Analysis and calibration of lombard effect and whisper for speaker recognition,''
\newblock {\em IEEE Transactions on Audio, Speech, and Language Processing}, vol. 29, pp. 927--942, Feb. 2021.

\bibitem{zhang2018advancements}
C.~Zhang, J.~H.~L. Hansen, and H.~A. Patil,
\newblock ``Advancements in whispered speech detection for interactive/speech systems,''
\newblock in {\em Signal Acoust. Model. Speech Commun. Disord.}, vol.~5, pp. 9--32. De Gruyter, 2018.

\bibitem{ZhangHansen2015WhisperIsland}
C.~Zhang and J.~H.~L. Hansen,
\newblock ``An advanced entropy-based feature with frame-level vocal effort likelihood space modeling for distant whisper-island detection,''
\newblock {\em Speech Communication}, vol. 66, pp. 107--117, 2015.

\bibitem{Alku2024AVID}
P.~Alku, M.~Kodali, M.~Laaksonen, and S.~R. Kadiri,
\newblock ``Avid: A speech database for machine learning studies on vocal intensity,''
\newblock {\em Speech Communication}, vol. 157, pp. 103039, 2024.

\bibitem{Kodali2025WSN}
M.~Kodali, S.~R. Kadiri, S.~Narayanan, and P.~Alku,
\newblock ``Wavelet scattering network features for intensity category classification and prediction of spl from speech,''
\newblock in {\em Proc. IEEE ICASSP}, 2025.

\bibitem{Baevski2020Wav2Vec2}
A.~Baevski, H.~Zhou, A.~Mohamed, and M.~Auli,
\newblock ``wav2vec 2.0: A framework for self-supervised learning of speech representations,''
\newblock in {\em Advances in Neural Information Processing Systems (NeurIPS)}, 2020, pp. 12449--12460.

\bibitem{Hsu2021HuBERT}
W.-N. Hsu, B.~Bolte, Y.-H.~H. Tsai, K.~Lakhotia, R.~Salakhutdinov, and A.~Mohamed,
\newblock ``Hubert: Self-supervised speech representation learning by masked prediction of hidden units,''
\newblock {\em IEEE/ACM Transactions on Audio, Speech, and Language Processing}, vol. 29, no. 9, pp. 3451--3460, 2021.

\bibitem{Gong2021AST}
Y.~Gong, Y.-A. Chung, and J.~Glass,
\newblock ``Ast: Audio spectrogram transformer,''
\newblock in {\em Proc. Interspeech}, 2021, pp. 571--575.

\bibitem{Kodali2024FineTuning}
M.~Kodali, S.~R. Kadiri, and P.~Alku,
\newblock ``Fine-tuning of pre-trained models for classification of vocal intensity category from speech signals,''
\newblock in {\em Proc. Interspeech}, 2024, pp. 482--486.

\bibitem{Kodali2023Whisper}
M.~Kodali, S.~R. Kadiri, and P.~Alku,
\newblock ``Comparison of ssl embeddings for vocal intensity classification in non-calibrated conditions,''
\newblock in {\em Proc. IEEE ICASSP}, 2023, pp. 1--5.

\bibitem{Omidi2024EndToEndHDSR}
Z.~Omidi and B.~Babaali,
\newblock ``On usage of an end-to-end deep neural architecture for handwritten digit string recognition,''
\newblock {\em Signal, Image and Video Processing}, vol. 18, no. 4, pp. 3009--3020, 2024.

\bibitem{Park2019SpecAugment}
D.~S.~Park et~al.,
\newblock ``Specaugment: A simple data augmentation method for automatic speech recognition,''
\newblock in {\em Proc. Interspeech}, 2019, pp. 2613--2617.

\bibitem{Kim2021SpecMix}
G.~Kim, D.~K. Han, and H.~Ko,
\newblock ``Specmix: A mixed sample data augmentation method for training with time-frequency domain features,''
\newblock in {\em Proc. Interspeech}, 2021, pp. 388--392.

\bibitem{Muller2019LabelSmoothing}
R.~Müller, S.~Kornblith, and G.~Hinton,
\newblock ``When does label smoothing help?,''
\newblock in {\em Advances in Neural Information Processing Systems (NeurIPS)}, 2019, vol.~32, pp. 4696--4705.

\bibitem{Xu2020SoftLabel}
J.~Xu and W.~Zhou,
\newblock ``Soft label training for deep neural networks,''
\newblock in {\em Proc. International Joint Conference on Neural Networks (IJCNN)}, 2020, pp. 1--8.

\bibitem{Zhang2018Mixup}
H.~Zhang, M.~Cisse, Y.~N. Dauphin, and D.~Lopez-Paz,
\newblock ``mixup: Beyond empirical risk minimization,''
\newblock in {\em Proc. International Conference on Learning Representations (ICLR)}, 2018.

\bibitem{Yun2019CutMix}
S.~Yun, D.~Han, S.~J. Oh, S.~Chun, J.~Choe, and Y.~Yoo,
\newblock ``Cutmix: Regularization strategy to train strong classifiers with localizable features,''
\newblock in {\em Proc. IEEE International Conference on Computer Vision (ICCV)}, 2019, pp. 6023--6032.

\bibitem{Chen2022WavLM}
S.~Chen et~al.,
\newblock ``Wavlm: Large-scale self-supervised pre-training for full stack speech processing,''
\newblock {\em IEEE Journal of Selected Topics in Signal Processing}, vol. 16, no. 6, pp. 1505--1518, 2022.

\bibitem{Hansen2024FearlessSteps}
J.~H. L.~Hansen et~al.,
\newblock ``Fearless steps apollo: Towards community resource development for science, technology, education, and historical preservation,''
\newblock in {\em Proc. IEEE Int. Conf. Acoust., Speech, Signal Process. (ICASSP)}, Seoul, Korea, Apr. 2024, pp. 12816--12820.

\end{thebibliography}

\end{document}